\documentclass[%
 reprint,
 amsmath,amssymb,
 aps,
]{revtex4-2}

\usepackage{graphicx}
\usepackage{dcolumn}
\usepackage{bm}

\begin{document}

\preprint{APS/123-QED}

\title{Axion-photon-mixing dark matter conversion mediated by torsion mass constrained by the Barbero-Immirzi parameter} 

\thanks{1. Xinjiang Astronomical Observatory, Chinese Academy of Sciences, Urumqi, Xinjiang, 830011,
China, zhifugao@xao.ac.cn; 2. Departamento de Fisica Teorica-IFADT-UERJ, Brasil and Institute for Cosmology and natural 
philosophy at Croatia.luizandra795@gmail.com}

\author{Zhi-Fu Gao$^1$ and Luiz C. Garcia de Andrade$^{2}$}


\begin{abstract}
In the Standard Model\,(SM) of particle physics, photon-torsion mixing is extended to include the Einstein-Cartan portal 
to dark-photon-axion-torsion mixing beyond the Standard Model\,(BSM), mediated by torsion. The Barbero-Immirzi(BI) 
parameter, of the order of $10^{-31}$, is more stringent than those obtained by Aliberti and Lambiase using 
matter-antimatter asymmetry. This paper presents the coupling of the SM with dark matter\,(DM) axions, both mediated 
by torsion. We discuss tordions, the quanta of torsion, and the damping of propagating torsion. It is shown that with 
both kinds of vectorial torsion masses, equations from Einstein-Cartan-Holst gravity can be derived, which reduce to 
axionic photon equations where torsion appears only through its mass spectrum. Photon-axion conversions and axion mixing 
are found to depend on the BI parameter. This study demonstrates that when the spin-0 torsion mass is finite and Proca 
electrodynamics is not ghost-free, dark axion masses align with spin-0 torsion masses via axion-driven torsion and 
photon-torsion mixing. Our results provide innovative insights into Proca gravity models and the role of torsion in 
photon-axion conversion and dark matter dynamics, thereby offering a solid foundation for future research and new 
theoretical frameworks in quantum gravity.
\end{abstract}
\keywords{Photon-torsion mixing, Barbero-Immirzi parameter, Nieh-Yan term, Axion-driven torsion, Dark matter dynamics}

\maketitle

\section{\label{sec:level1}Introduction}
In recent years, there has been a growing interest and activity \cite{1} in the Einstein-Cartan\,(EC) portal involving 
fermions and bosons\,\cite{2} from the Standard Model\,(SM) of particle physics, as well as Beyond-Standard-Model\,(BSM) 
dark matter\,(DM), one of the universe's mysteries. On the pathway of quantum gravity with torsion, other sources of active 
research include the investigation of ghosts and tachyon-free scenarios in the propagation of torsion, and tordions \cite{3} 
the quanta of torsion in Poincaré gauge gravity\cite{4}. Loop quantum gravity, another theory of quantum gravity, has also 
introduced an important free parameter known as the Barbero-Immirzi\,(BI) parameter\,\cite{5}. Barbero-Immirzi parameter 
 is a numerical coefficient appearing in loop quantum gravity\,(LQG), a nonperturbative theory of quantum gravity. 
It measures the size of the quantum of area in Planck units. These topics illuminate the complex subject of quantum gravity 
within alternative theories. Quantum torsion, sometimes referred to as tordions, has been explored by several authors\cite{6}
both in connection with propagating torsion waves and independently. From a mathematical perspective, the Holst and Nieh-Yan 
(NY) terms are topological invariants that involve the BI parameter. In this paper, we have integrated these topics into 
Kruglov's photon-torsion mixing\,\cite{7} as a mediation to the Einstein-Cartan\,(EC) portal for DM. They are represented by
\begin{equation}
{\cal{H}}= R_{ijkl}{\epsilon}^{ijkl},
\label{1}
\end{equation}
where ${\epsilon}^{ijkl}$ (with $i,j, k,l=0, 1, 2, 3$) is the totally-skew-symmetric Levi-Civita\,(LÇ) symbol, and $R$
 represents the four-dimensional Riemann-Cartan (RC) curvature and torsion tensor. The NY invariant is defined as
\begin{equation}
{\cal{NY}}= {\partial}_{i}(T_{jkl}{\epsilon}^{ijkl})={\partial}_{i}S^{i},
\label{2}
\end{equation}
where $T$ is the Cartan torsion tensor with three indices\,\cite{8}, and $S$ is the totally 
symmetric part of the torsion tensor. In this work the usual $\sqrt{-g}$($g=det(g_{ik})$ is the metric tensor determinant) 
is omitted in the above formulas due to its minimal impact on torsion at LHC CERN hadron accelerators where the effect of 
Riemannian curvature is minimum\,\cite{4}. Explicitly, the totally skew-symmetric part of the RC curvature tensor in a 
spacetime manifold endowed with torsion and curvature is given by
\begin{equation}
{\cal{H}}\supset{T_{k}{S}^{k}},
\label{3}
\end{equation}
where the symbol $\supset$ means ``contains", indicating that $\cal{H}$ might include other terms or factors.
In this Holst term, we notice the presence of a mixing term between two types of torsion vectors and pseudo-vectors.
The RC Ricci tensor is expressed as
\begin{equation}
R={\tilde{R}} -2{\nabla}_{i}T^{i}-\frac{2}{3}T^{i}T_{i}+\frac{1}{24}S_{i}S^{i}.
\label{4}
\end{equation}
On the right side of the equation, the first term ${\tilde{R}}$ represents the Ricci-Riemannian scalar,which accounts solely 
for the curvature part in Riemannian geometry, excluding any contributions from torsion. The second term signifies 
the divergence of the torsion vector $T^{i}$, with $\nabla_{i}$ denoting the covariant derivative. The third term captures 
the self-interaction of the torsion vector $T^{i}$, highlighting the complexities of torsion dynamics. Finally, the 
fourth term encapsulates the self-interaction of the pseudovector component of the torsion tensor $S_{i}$, adding another 
layer of interaction within the torsion framework.

In this paper, we utilize the EC theory, a minimal and widely accepted alternative to General Relativity\,(GR). This theory, 
also known as Einstein-Cartan-Kibble-Sciama gravity \cite{9}, incorporates non-dynamical torsion to address spin-spin 
four-fermion interactions, which do not propagate in a vacuum. Conversely, its dynamical torsion propagates in the form of 
torsion waves. Previously, Duncan et al. \cite{10} explored the coupling of axions with torsion by considering the axial 
pseudo-torsion vector. This coupling has been investigated as DM axion gradients in galaxies\,\cite{11}. Here, we 
extend this concept by coupling torsion-induced dark axions with Kruglov's model\,\cite{7}, where propagating torsion 
interacts with electromagnetic SM photons. This demonstrates the potential for coupling dark photons and examining their 
abundance\,\cite{12} through standard massless Maxwell photons. Consequently, EC gravity acts as a portal between SM and 
BSM dark matter.

Shaposhnikov et al.\,\cite{1} have studied the EC portal to DM \cite{13, 14}, which is not predicted by particle physics 
DM, and have explored its implications in quantum gravity \cite{15}. Using the EC portal, researchers such as C. H. Cao 
\cite{16} have addressed gauge dark bosons, while Elgammal et al.\cite{17} have investigated gauge dark bosons and 
fermionic DM. Torsion mediates the coupling between SM and BSM via the EC portal. The presence of the NY term in the action 
is fundamental for introducing the BI parameter in the propagating torsion in this work. Theoretically, Barker et al.\cite{18} have studied the role of the torsion trace vector in 
metric-affine gravity, where Weyl non-metricity can also be present. In their work, dynamical torsion introduces ghosts 
that can only be removed by considering scalar torsion \cite{19}. Recently, Barker and Zell \cite{20} have identified 
additional challenges in propagating torsion. Here we show that ghosts and tachyons may appear in Cartan-Proca-Holst-
Nieh-Yan photon-torsion mixing with  respective topological terms. In this Cartan-Proca gravity framework, it is shown 
that boson quantization can eliminate ghosts and tachyons as the BI parameter approaches infinity. The DM sector can 
emerge in these theories by obtaining an expression where ghosts and tachyons are removed. From the cosmological dark 
sector for bouncing cosmology, it is possible to see that the mixed Holst term of both torsion modes may yield axion-torsion 
as a candidate for DM.

The remainder of this paper is organized as follows: Section 2 tackles the complexities of quantum corrections and ghost 
problems in axion-driven torsion within the Einstein-Cartan-Proca\,(ECP) theory, deriving constraints on the BI parameter 
and examining torsion-electromagnetic field interactions to provide an in-depth analysis of the dynamics and implications 
in Proca gravity. Section 3 delves into determining the BI parameter in the context of axion-driven torsion within loop 
quantum gravity, highlighting its role and significance. Section 4 establishes the derivation of torsion-photon coupling 
from the electromagnetic\,(EM) equation, demonstrating its dependence on constant torsion divergence and exploring profound 
implications for dark matter dynamics within the Proca-magnetogenesis framework. Section 5 identifies and eliminates 
tachyons and ghosts through Fourier transform applications to Proca and torsion propagation equations, resulting in a 
robust, ghost-free, and tachyon-free model suitable for quantum gravity with torsion. Section 6 illustrates that, although 
the theory is not initially ghost-free, it becomes so as the BI parameter approaches infinity, aligning with classical 
EC theory. Section 7 explores the pivotal role of Einstein-Cartan-Holst gravity in photon-axion conversion, 
particularly considering axion-torsion transmutation. Lastly, Section 8 summarizes our findings and provides a comprehensive 
outlook on future research directions, emphasizing the potential breakthroughs and advancements our work facilitates.

\section{Dynamics of Torsion Propagation in Proca-Nieh-Yan Gravity}
In this section, we address the challenges posed by quantum corrections and ghost problems arising from axion-driven torsion, 
which need to be resolved. To restore the quantum gravity torsion formulation of the Einstein-Cartan-Proca\,(ECP) theory, even in the absence of 
quadratic gravity terms, we first express the Lagrangian for this model
\begin{eqnarray}
{{{\cal{L}}_{ECP}}\supset = {\frac{1}{2}m^{2}_{s}S^{2}-{\gamma}\frac{m^{2}_{P}}{2}({\partial}{S})}}\nonumber \\ {-\frac{1}{2}m^{2}_{P}({\partial}_{[i}S_{j]})^{2}+C},
\label{5}    
\end{eqnarray}

where $\frac{1}{2}m^{2}_{s}S^{2}$ is the mass term for the torsion field $S$, signifying the self-interaction energy 
related to mass $m_s$. The term $-{\gamma}\frac{m^{2}_{P}}{2}({\partial}{S})$ accounts for the interaction between the 
field $S$ and its derivatives, modulated by the Immirzi parameter $\gamma$, The kinetic term $-\frac{1}{2}m^{2}_{P}
({\partial}_{[i}S_{j]})^{2}$ represents the dynamic energy of the torsion field $S$, impacting its propagation in spacetime. 
The auxiliary function $C$ is given by
\begin{equation}
C(b,S,T)=b({\partial}S)^{2}-{\gamma}m^{2}_{P}TS+\frac{1}{2}m^{2}_{T}T^{2}.
\label{6}
\end{equation}
where $\gamma$ is the inverse of the IB parameter $\beta$, $m_{P}=10^{19}$\,GeV is the Planck mass, 
$m_s$ and $m_{T}$ represent the masses associated with torsion, axial, and vector modes, $S=S_{i}$, ${\partial}{S}=
{\partial}_{k}{S}^{k}$ and $({\partial}_{[i}S_{j]})^{2}$ is the square of the gravitational term. The parameter $b$
 represents the quantum correction term, with indices $i=0,1,2,3$. 

To build a quantum Einstein-Cartan-Proca gravity, and eliminate ghosts, we set $b=0$, which simplifies the Lagrangian to
\begin{equation}
{{\cal{L}}_{ECP}} \supset{\frac{1}{2}m^{2}_{s}S^{2}-{\gamma}\frac{m^{2}_{P}}{2}({\partial}{S})+A(S,T,T^{2})},
\label{7}
\end{equation}
where
\begin{equation}
A(S,T,T^{2})= -\frac{1}{2}m^{2}_{P}({\partial}_{[i}S_{j]})^{2}-{\gamma}m^{2}_{P}TS+\frac{1}{2}m^{2}_{T}T^{2}.
\label{8}
\end{equation}
To derive the necessary constraints, we perform the variation of this Lagrangian, incorporating the Holst and Nieh-Yan 
terms, with respect to $\delta{T}$, resulting in
\begin{equation}
{\gamma}\frac{m^{2}_{P}}{m^{2}_{T}}S=T.
\label{9}
\end{equation}
Substituting this constraint into Equation (\ref{4}) and varying the Lagrangian with respect to $S$ yields
\begin{equation}
\frac{m^{4}_{P}}{{\beta}^{2}m^{2}_{T}}S-\frac{m^{2}_{P}}{2}{\Box}{S}-\frac{1}{\beta}m^{2}_{P}\partial({\partial}S)+
m^{2}_{T}S=0,
\label{10} 
\end{equation}
utilizing the relation $\beta=1/\gamma$. To solve this torsion wave equation, decoupling it from the 
dark photon and torsion mixing (addressed in the next section), we assume the ansatz
\begin{equation}
S=S_{0}exp[i({\omega}t-kz)],
\label{11}
\end{equation}
assuming the torsion spin-1 wave propagation along the $z$-direction. Notably, torsion propagates outside matter, unlike 
non-dynamical original E-C gravity\,\cite{21}. In contrast to Kruglov \cite{7}, no external magnetic field is required. 
Generalizing Kruglov's work to Einstein-Cartan-Proca gravity, substituting Equation (\ref{6}) into Equation (\ref{5}) gives
\begin{equation}
{\omega}^{2}+({\gamma}-1){k}^{2}-{\gamma}k{\omega}-c_{0}{\gamma}^{2}=0,
\label{12}
\end{equation}
and 
\begin{equation}
({\gamma}-1){\omega}^{2}+{\gamma}k{\omega}-c_{0}{\gamma}^{2}=0,
\label{13}
\end{equation}
where $c_{0}= \frac{m^{4}_{P}}{m^{2}_{T}}$. These two equations are obtained respectively by taking $i=1$ and 
$i=z=3$. Combining them leads to an algebraic characteristic eigenvalue equation for the IB parameter
\begin{equation}
-4c_{0}{\gamma}^{2}+{\gamma}-k^{2}+{\gamma}k^{2}={0}.
\label{14}
\end{equation}
Assuming resonant BI parameter ${\gamma}_{+}={\gamma}_{-}$, implying a vanishing discriminant 
$\Delta$, the general solution is
\begin{equation}
{\gamma}= \frac{({\omega}^{2}+k^{2}){\pm}\sqrt{\Delta}}{c_{0}},
\label{15}
\end{equation}
with  
\begin{equation}
{\Delta}= \frac{({\omega}^{2}+k^{2})^{2}-8k{4\sqrt{c_{0}}}}{c_{0}}.
\label{16}
\end{equation}
Solving  for ${\omega}^{2}$ and substituting into Equation (\ref{11}) gives the BI parameter in terms of the torsion 
wave vector of the torsion wave
\begin{equation}
\gamma=2\frac{m_{T}}{m^{2}_{P}}k.
\label{17}
\end{equation}
Taking $k=\frac{1}{\lambda}$ and considering the length scale $l_{P}=10^{-33}$\,cm, transformed to GeV units, we obtain
the denominator is the length scale in the universe where we are measuring 
\begin{equation}
{\gamma}={2}{\times}10^{-32}.
\label{18}
\end{equation}
This matches the value obtained by Aliberti and Lambiase \cite{22} for matter-anti-matter asymmetry. The following table 
presents data for the BI parameter across various cosmic scales, converted to GeV units:
\begin{table}[h!]
\centering
\begin{tabular}{ccc}
\hline
               Einstein-Cartan-Proca   &$\lambda$\,cm &   BI(LQG)\\
\hline
               Planck scale           &$10^{-33}$       &$10^{-32}$   \\
               Cosmic Dust             &$10^{-6}$        &$10^{-43}$  \\
               Nuclei                  &$10^{-8}$       &$10^{-41}$   \\
\hline
\end{tabular} \caption{BI parameter from distinct cosmic eras in the universe}
\label{table:formulas}
\end{table}
All computations were performed in GeV units. In the next section, we will calculate the BI parameter based on another 
physical axial anomaly involving electromagnetic fields and compare the results.

\section{BI Parameter Boubds Induced by Dark-Photons and Axio-Torsion  Mixing}
Motivated by the work of Lazanttani and Mercuri\,\cite{23} on the BI field as an axion itself and the solution to 
the strong CP problem via torsion with the Nieh-Yan (NY) term, as well as the work of Karananas \cite{24}, we 
have recently investigated QCD axion-torsion couplings, exploring light torsion as a dark matter candidate. Building 
on these foundations, this section focuses on determining the LQG BI parameter in the context of axion-driven 
torsion. By incorporating the NY term into the Lagrangian\,(\ref{21}), we derive the following
\begin{equation}
{\cal{L}}_{ECP}\supset{{\frac{1}{2}m^{2}_{s}S^{2}-{\gamma}\frac{m^{2}_{P}}{2}(\partial{S})-{\gamma}m^{2}_{P}TS+
D(T,{\phi},F^{2})}},
\label{19}
\end{equation}
where 
\begin{equation}
D(T,{\phi},F^{2})=\frac{1}{2}m^{2}_{T}T^{2}+{\phi}F\tilde{F}-\frac{1}{4}F^{2}-\frac{1}{2}m^{2}_{\phi}{\phi}^{2}.
\label{20}
\end{equation}
In this Lagrangian, the squared gravity term is absent as $S_{[i,j]}$ vanishes due to Minkowskian nature of BI 
torsion, being the gradient of the dark axion. Holst and NY terms are present and $b=0$ ensures a ghost-free theory. By 
substituting $S$, an axial pseudo-vector torsion, with the gradient of a Nambu-Goldstone pseudo-scalar axion boson $\phi$,
our Cartan-Proca Lagrangian becomes a pure scalar Lagrangian where the axion is fundamental. Varying this equation with 
respect to the axion yields 
\begin{equation}
{m^{2}}_{s}[1+({\gamma}^{2}-{\gamma})\frac{m^{4}_{P}}{m^{2}_{s}m^{2}_{T}}]{\Box}{\phi}+m_{\phi}^{2}{\phi}=
{\textbf{E}}\cdot\textbf{B}.
\label{21}
\end{equation}
To simplify matters, we find the BI parameter in the case of an anomalous chiral source term
\begin{equation}
{\textbf{E}}\cdot{{\textbf{B}}}={0},
\label{22}
\end{equation}
assuming orthogonal electric and magnetic fields in the generated electromagnetic wave. Considering the second term 
in the brackets as much weaker than one, we derive an upper bound for the BI parameter:
\begin{equation}
{\gamma}\ll 10^{-52}.
\label{23}
\end{equation}
This bound is extremely stringent and not in agreement with other previous bounds, although not significantly different. 
This suggests that, as far as the BI parameter is concerned, we might favor the Proca formulation with the Holst term as 
discussed in Section 2. Next, we generalize the photon-torsion mixing proposal. In Proca gravity, as presented by Kruglov, 
the axion-driven torsion formulation naturally introduces the Proca equation for scalar fields, which minimally coincides 
with the Klein-Gordon equation. Lut us reproduce Kruglov's interaction Lagrangian
\begin{equation}
{\cal{L}}_{int}=\frac{1}{4}g({\partial}_{k}S^{k})F{\tilde{F}},
\label{24}
\end{equation}
where $\tilde{F}$ is the dual of the EM tensor represented by $F=F_{ij}$. Utilizing the axion-driven torsion of Duncan et 
al.\cite{10}, not addressed by Kruglov, this interaction Lagrangian becomes
\begin{equation}
{\cal{L}}_{int}=\frac{1}{4}g({\Box}{\phi})F{\tilde{F}},
\label{25}
\end{equation}
with electromagnetic equations become
\begin{equation}
{\partial}_{i}F^{ij}-g{\partial}_{i}({\Box}{\phi}){\tilde{F}}^{ij}=0.
\label{26}
\end{equation}
These equations can be solved, or a dynamo equation can be derived by simply substituting the D'Alembertian wave 
operator with an axion function. Therefore, obtaining the axion equation for DM requires substituting the 4-gradient 
Minkowskian operator of axion-driven torsion coupling into the torsion wave equation
\begin{equation}
{\Box}S_{i}-\delta{\partial}_{i}({\partial}_{k}S^{k})- m^{2}S_{i}=\frac{1}{4}{\partial}_{i}(F{\tilde{F}}),
\label{27}
\end{equation}
with constant 
\begin{equation}
{\delta}=[1-\frac{m^{2}}{m_{0}^{2}}].
\label{28}
\end{equation}
Here $m$ is the spin-1 torsion mass, and $m_{0}$ is the spin-0 torsion mode. It is interesting that, for a ghost-free axion-
driven torsion Proca gravity formulation, $m_{0}$ must approach infinite, making $\delta=1$, and reducing the equation to 
\begin{equation}
{\partial}_{i}({\Box}\phi)-{\delta}{\partial}_{i}({\Box}\phi)-{m^{2}}S_{i}=\frac{1}{4}{\partial}_{i}
(\textbf{E}\cdot{\textbf{B}}).
\label{29}
\end{equation}
Considering the non-chiral or non-axial anomaly of the EM field with orthogonal electric and magnetic fields, 
we obtain the equation for the axion,
\begin{equation}
{\Box}{\phi}+{m^{2}}_{0}{\phi}=0.
\label{30}
\end{equation}
This shows the spin-0 torsion mass coincides with the axion mass in Proca electrodynamics with photon-torsion-axion 
mixing. Now, adding the Holst term $TS$ multiplied by the BI parameter into the photon-torsion mixing interaction, 
we obtain
 \begin{equation}
{\cal{L}}_{int}=\frac{1}{4}g({\partial}_{k}S^{k})F{\tilde{F}}+m^{2}_{P}{\gamma}T_{k}S^{k}.
\label{31}
\end{equation}
Using the same axion torsion-driven and photon-torsion mixing assumption that the torsion vector mass vanishes, we derive
\begin{equation}
{m^{2}}_{s}[\frac{\gamma}{m^{2}_{P}}-\frac{1}{m^{2}_{0}}]{\Box}{\phi}+m^{2}_{\phi}{\phi}=0.
\label{32}
\end{equation}
From this equation, the value of the BI parameter ($\beta=\frac{1}{\gamma}$) is estimated as 
\begin{equation}
\beta \gg \frac{m^{2}_{s}}{m^{2}_{P}}=10^{-52},
\label{33}
\end{equation}
 yielding
\begin{equation}
{\beta} \gg 10^{-26}.
\label{34}
\end{equation}
This aligns with the BI parameter induced by matter-antimatter asymmetry as discussed by Aliberti and Lambiase. 
To obtain this upper bound for the BI parameter, we assume the spin-0 torsion mass is equal to the axion 
mass, $m^{2}_{0}={m^{2}_{s}}$. Now let's explore magnetogenesis by solving the Proca electrodynamic equation 
\begin{equation}
{\nabla}\cdot{\textbf{E}}+\frac{gm^{2}_{\phi}}{4}{\phi}[\textbf{k}\cdot\textbf{B}]=0.
\label{35}
\end{equation}
This is analogous to the Coulomb-Maxwell equation in photon-torsion Proca context. Here, "Proca equation" refers to 
massive torsion and axions, not just the Proca massive photon. To derive this last equation, we have used 
the no-monopole equation, $divB=0$. The remaining equations are
\begin{equation}
{\partial}_{t}{\textbf{B}}=- \nabla \times (\phi \textbf{E}).
\label{36}
\end{equation}
A simple solution for this Faraday equation is given by
\begin{equation}
{B^{z}}={B}_{0}e^{{\omega}_{1}t-k_{z}z},
\label{37}
\end{equation}
where the resonance ${\omega}_{1}=\omega$ was used, and $B_{0}=-i{\phi}_{0}k$. This results in an oscillating magnetic 
field induced by the axion-torsion-driven field in photon-torsion mixing with spin-0 and spin-1 torsions. Note that in 
these examples, the vector torsion has also been used as discussed by Barker and Zell\,\cite{18}.

\section{Photon-Torsion Coupling in Proca Dark Magnetogenesis Sector}
In this section, we demonstrate the possibility of deriving the torsion-photon coupling expression in terms of a 
constant torsion divergence and wave equation from the EM equation
\begin{equation}
[{\partial}_{t}^{2}-{\nabla}^{2}]\textbf{A}-c_{0}g[\lambda\textbf{A}+2{\lambda}_{0}
{{\partial}_{t}}\textbf{A}]=0,
\label{38}
\end{equation}
where
\begin{equation}
{\nabla}{\times}\textbf{A}=\lambda{\textbf{A}}.
\label{39}
\end{equation}
This describes magnetic helicity, applicable to EM fields. The dispersion 
relation is given by:
\begin{equation}
{\omega}^{2}+2gic_{0}{\lambda}^{2}{\omega}+(k^{2}+c_{0}g{\lambda}^{2})=0.
\label{40}
\end{equation}
We solve this equation under resonance conditions where the discriminant ${\Delta}$ vanishes. From a very low-frequency 
EM wave or very low-frequency\,(VLF) suitable for detection in the VLF array, we obtain
\begin{equation}
 g=- c_{0}.
\label{41}
\end{equation}
Therefore, the photon-torsion coupling is dependent on a constant torsion divergence. In this scenario, only the EM wave 
survives, and torsion waves are absent in this Proca regime. Next, we explore the potential for DM dynamos in the dark 
Proca-magnetogenesis sector by solving the above equation in a more general frame where the 4-divergence of the axial 
torsion vector does not vanish. This more complex solution for non-chiral magnetic fields is described by the field 
equations
\begin{equation}
\frac{d^{2}}{d{\eta}^{2}}\textbf{A}-g\lambda[\frac{d{S}_{0}}{d{\eta}}({\lambda}+im_{0})+{\frac{d^{2}S_{0}}{d{\eta}}^{2}}]
{\textbf{A}}=0
\label{42}
\end{equation}
where $\eta$ is the conformal spacetime coordinate. Let's take the ansatz
\begin{equation}
{S}_{0}=exp[i{\omega}{\eta}].
\label{43}
\end{equation}
At the early universe, ${\eta}\rightarrow{0}$, and we approximate 
\begin{equation}
{S}_{0}\approx{i{\omega}{\eta}},
\label{44}
\end{equation}
where  
\begin{equation}
{\omega}^{2}=g{\lambda}{m_{0}}^{2}
\label{45}
\end{equation}
and the electromagnetic solution is 
\begin{equation}
\textbf{A}= {\textbf{A}}_{0}exp(i\omega{\eta}).
\label{46}
\end{equation}
Note that to achieve this solution, we use the result that the axion mass and torsion mass approximately coincide. 
From expression (\ref{41}), we notice that the magnetic helicity can be expressed in terms of the photon-torsion mixing 
constant $g$ and the spin-0 mass. Assuming that torsion propagates at UHF of 1 GHz, we have
\begin{equation}
\lambda= \frac{\omega}{m_{0}g}\approx{10^{-12}},
\label{47}
\end{equation}
using a spin-0 mass of 1 TeV and $g=10^{-24}$ for a massive photon. At this TeV scale, the magnetic helicity is minimal, 
indicating weak coupling with massive photons, which could be potential dark matter candidates. Note that the magnetic 
field can be obtained using the relation $B=iKA$. The oscillation frequency of the magnetic field depends on the magnetic 
helicity torsion-photon coupling $g$ and the spin-0 axion mass. In the next section, we explore hypotheses on the tachyonic 
Proca equation and its ghosts within the Proca theory framework.

\section{Ghosts and Tachyons in Fourier Spectrum of Photon-Torsion Mixing Self-Dual Proca Fields}
Previously, Barker and Zell\,\cite{18} showed that tachyons and ghosts appearing in Proca equations in Riemann-Cartan 
spacetime could be resolved by applying constraints on torsion. In this section, we extend this analysis by 
applying the Fourier transform $ik^{j}\rightarrow{{\partial}^{j}}$ to the Proca and torsion propagation equations, 
aiming to identify the potential for finding tachyons and ghosts. We begin by expressing these equations with this transform:
\begin{equation}
ik_{l}F^{lp}+gk_{m}k_{l}S^{l}{\epsilon}^{mprs}F_{rs}=0,
\label{48}
\end{equation}
where the dual of the EM field tensor is 
\begin{equation}
{\epsilon}^{mprs}F_{rs}={\tilde{F}}^{mp}.
\label{49}
\end{equation}
By taking the complex dual of $F$ as $iF_{rs}={\tilde{F}}_{rs}$, expression (\ref{49}) simplifies to
\begin{equation}
iF^{mp}=-gk_{l}S^{l}{\tilde{F}}_{rs}.
\label{50}
\end{equation}
Applying the self-dual relation, we obtain
\begin{equation}
g^{-1}=-k_{l}S^{l}> 0.
\label{51}
\end{equation}
This implies that if the EM wave vector is light-like, $k^{s}k_{s}=-m^{2}<0$, as we shall see below, 
it is space-like or tachyonic. Meanwhile, the torsion wave expression becomes
\begin{equation}
-(k_{s}^{2}+m^{2}_{s})+{\delta}S^{i}(k^{l}k_{l})^{s}=0.
\label{52}
\end{equation}
An important issue in this equation is that $k$, with indices indicating axial torsion, is not a 
null vector, unlike the EM wave vector where
\begin{equation}
{k}^{i}k_{i}=0.
\label{53}
\end{equation}
From the first set (\ref{52}), one obtains
\begin{equation}
k_{i}F^{ij}=0.
\label{54}
\end{equation}
From the conformal flat metric:
\begin{equation}
ds^{s}=d{\eta}^{2}-d{\textbf{x}}^{2},
\label{55}
\end{equation}
the expression (\ref{47}) reads
\begin{equation}
({{\partial}_{\eta}}^{2}+{m^{2}}_{s})S^{i}+{\delta}S^{i}(k^{l}k_{l})^{s}=0.
\label{56}
\end{equation}
Multiplying the last equation by torsion wave vector $k^{s}$, we get
\begin{equation}
[-({k}_{s}^{2}+m^{2}_{s})+{\delta}(k^{l}k_{l})^{s}]\alpha=0,
\label{57}
\end{equation}
where $\alpha= k^{s}\cdot S$. Thus, we obtain
\begin{equation}
({\partial}_{\eta}^{2}+m^{2}_{s})+{\delta}(k_{s})^{2}=0.
\label{58}
\end{equation}
There are two cases to consider: the ghost-free case $\delta=1$
\begin{equation}
[(-{k}_{s}^{2}+m^{2}_{s})+(k^{l}k_{l})^{s}]\alpha=0,
\label{59}
\end{equation}
which reduces to the vanishing of torsion mass. Next, we examine the case of a finite ghost mass $m_{0}$, yielding
\begin{equation}
-{k}_{s}^{2}+m^{2}_{s})=0,
\label{60}
\end{equation}
indicting a relation $(k^{l}k_{l})^{s}>0$, meaning torsion vector propagation is not tachyonic. Therefore, the Proca 
equation with spin-0 and spin-1 states are tachyonic-free. Assuming it is also ghost-free with an axial anomaly
\begin{equation}
({\partial}_{\eta}^{2}+m^{2}_{s})S^{i}+k_{s}^{2}S^{i}={\partial}^{i}(F\tilde{F}).
\label{61}
\end{equation}
Using the self-dual identity $iF=\tilde{F}$ into the RHS of the last equation, we find that the source is identical 
to $F^{2}=(E^{2}-B^{2})$. Substituting these expressions into the previous formula, we get
\begin{equation}
[({k}_{s}^{2}+m^{2}_{s})+(k_{s})^{2}]k^{i}_{i}S^{i}=({k}^{i}_{s})^{2}(B^{2}-E^{2}),
\label{62}
\end{equation}
which yields
\begin{equation}
m^{2}_{s}k^{i}_{i}S^{i}=({k}^{i}_{s})^{2}(E^{2}-B^{2}).
\label{63}
\end{equation}
Note that if ${k^{i}}_{i}S^{i} > 0$, and $B^{2}>E^{2}$, then ${{k}_{s}}^{2}>0$ and the torsion wave vector $k_{s}$ is 
time-like. Therefore, Proca-Cartan is tachyonic. It is possible to find a tachyonic-free, ghost-free Proca theory 
with photon-torsion mixing, which is free of pathologies in quantum gravity with torsion.

Next, we investigate what happens when $m_{0}$ approaches infinity, making Proca-Cartan ghost-free. 
The existence of tachyons is derived from the equation:
\begin{equation}
\frac{m^{2}_{0}-k^{2}_{s}}{m^{2}_{0}}k_{s}S= 
(\frac{m_{0}}{m_{s}})^{2}({{k}^{i}}_{s})^{2}(B^{2}-E^{2}).
\label{64}
\end{equation}
Notice that in this formula, when the Proca is ghost-free or $m_{0}$ goes to infinity, the mass coefficient tends to 
one on the LHS of the equation which becomes
\begin{equation}
k_{s}S= (\frac{m_{0}}{m_{s}})^{2}({k}^{i}_{s})^{2}(B^{2}-E^{2}),
\label{65}
\end{equation}
and if $k_{s}S>0$ and the magnetic energy density is higher than the electric one, we conclude that $({k}^{i}_{s})^{2}>0$,
indicating that the torsion wave vector $k_s$ is time-like. This demonstrates that the model is free of 
ghosts and tachyons, making it suitable for quantum gravity with torsion.

\section{Boson Quantization and Ghost-free Cartan-Proca Fields by Constraining BI Parameter}
Kruglov\,\cite{7} has shown that boson quantization applied to the propagating torsion equation leads to a ghost mass of 
spin-0 torsion, causing formulation problems. In the absence of the BI parameter and without Kruglov's photon-torsion 
mixing, Barker and Zell\,\cite{18} have already identified this problem in torsion propagation. In this section, we use the 
Lagrangian with Holst and NY terms to show that the propagator matrix can be expressed in terms of the BI 
parameter. Although this approach is not inherently ghost-free, in the limit where the BI parameter approaches infinity, 
the Cartan-Proca equation becomes ghost-free. This phenomenon is well known, even in classical physics, where in 
this BI limit, Einstein-Cartan-Holst-Nieh-Yan gravity reduces to the original EC theory. Moreover, this is analogous 
to Kruglov's photon-torsion mixing, where in the limit of spin-0 mass approaches infinity, the ghost of Proca theory with 
propagating torsion is eliminated. 

To demonstrate this, we express the equation in the form
\begin{equation}
-\frac{m^{2}_{P}}{{\beta}^{2}m^{2}_{T}}S+{\Box}S+{\beta}^{-1}\partial({\partial}S)-\frac{m^{2}_{T}}{m^{2}_{P}}S=0.
\label{66}
\end{equation}
By computing the $4x4$ matrix $M_{ij}$ used to determine the propagator, we have
\begin{equation}
M_{ik}=(p^{2}+m^{2}_{s}){\delta}_{ik}-2{\beta}^{-1}p_{i}p_{k}.
\label{67}
\end{equation}
This straightforward expression involving the four-momenta $p$ highlights that the second term has the wrong sign, 
indicating that the theory is not free of ghosts. Therefore, the only viable solution is to consider the scenario 
where the BI parameter approaches infinity, thereby achieving a ghost-free solution.

\section{Photon-Axion Conversion with Background of Torsion Mass Spectrum and FRB Constraints 
from BI Parameter}
In this section, by utilizing Duncan et al.'s axion-torsion transmutation\,\cite{10}, we show that Einstein-Cartan-Holst 
gravity can investigate the role played by the torsion mass spectrum on photon-axion conversion. When the BI parameter 
assumes a certain value, the axion-photon conversion equations are free from the torsion spectrum. We propose the 
following Lagrangian
\begin{eqnarray}
{\cal{L}}_{ECH}\supset{\frac{1}{2}({\partial}a)^{2}+{\frac{1}{2}m^{2}_{s}S^{2}
-{\gamma}\frac{m^{2}_{P}}{2}(\partial{a})T+D(T, a,F^{2})}}
\label{68}
\end{eqnarray}
 by following Duncan et al's proposal, the axion $a(z.t)$ is transmuted to torsion using the relation 
$S={\partial}a$. Here, $\partial$ denotes a suitable partial gradient derivation. The $z$-direction is chosen 
to align with the polarization direction. Before explicitly constructing the field equations and their perturbations 
to analyze the photon-axion-torsion conversion, we must express the auxiliary relations as follows
\begin{equation}
D(T,a,F^{2})=\frac{1}{2}m^{2}_{T}T^{2}+g_{a{\gamma}{\gamma}}aF\tilde{F}-\frac{1}{4}F^{2}-\frac{1}{2}m^{2}_{a}{a}^{2},
\label{69}
\end{equation}
where $g_{a{\gamma}{\gamma}}$ represents the axion-photon coupling constant.

In this section, we examine the influence of the Holst term on the torsion mass spectrum during 
photon-axion conversion within a constant magnetic field, denoted as ${\textbf{B}}_{0}=(B_{0},0,0)$. 
Initially, we demonstrate that varying the axion $a$ and the torsion trace vector $T$ yields significant results: 
the torsion trace vector $T$ depends on the gradient of $a$ and through appropriate substitutions, only the 
presence of torsion masses remains in the final field equations to be perturbed. Variation of this equation with 
respect to the axion results in
\begin{equation}
{\Box}a -{\gamma}{\partial}Tm^{2}_{P}+m^{2}_{a}a= g_{a\gamma\gamma}\textbf{E}\cdot\textbf{B}.
\label{70}
\end{equation}
To simplify, we assume the axial torsion mass vanishes in this section. By varying $T$ in the last Lagrangian, we get
\begin{equation}
T=\gamma\frac{m^{2}_{P}}{m^{2}_{T}}{\partial}a.
\label{71}
\end{equation}
This equation shows that the torsion trace vector depends on the gradient of the axion through a different expression. 
Substituting this expression into Equation (\ref{70}), we obtain
\begin{equation}
{\Box}a-{m^{2}_{a,eff}}=-\frac{1}{2}g_{a{\gamma}{\gamma}}{(\frac{m_{a,eff}}{m_{a}})}^{2}{\textbf{E}\cdot\textbf{B}}.
\label{72}
\end{equation}
where $m_{a,eff}$ denotes the effective axion mass. To obtain the effective axion mass. which appears in terms of the BI 
parameter, we assume the LHC result that torsion masses are very high, around 1\,TeV. The effective axion torsion mass 
is given by
\begin{equation}
m_{a,eff}=m_{a}{\beta}\frac{m_{T}}{m_{P}}.
\label{73}
\end{equation}
Now, we propose the following perturbations for axions and EM fields to build the Schrödinger-like 
matrix for axion-photon mixing and conversion. The perturbations are
\begin{equation}
a(z,t)=0+{\delta}a(z,t)
\label{74}
\end{equation}
and
\begin{equation}
\textbf{B}={\textbf{B}}_{0}+{\delta}\textbf{B}(z,t).
\label{75}
\end{equation}
Due to polarization, the constant magnetic field in modulus is denoted by ${\textbf{B}}_{0}=(B_{0},0,0)$. The 
perturbation in the electric field is given by
\begin{equation}
{\textbf{E}}=0+{\delta}\textbf{E}(z,t)
\label{76}
\end{equation}
which gives rise to the following wave equations
\begin{equation}
\Box\delta{A_{||}}=g_{a{\gamma}{\gamma}}B_{0}({\partial}_{t}{\delta}a)
\label{77}
\end{equation}
and 
\begin{equation}
\Box\delta{A_{ort}}=0.
\label{78}
\end{equation}
Combining this with the axion electrodynamics equation
\begin{equation}
{\partial}_{i}F^{ij}+F^{ij}{\partial}_{i}a=0,
\label{79}
\end{equation}
we derive the perturbative axion equation
\begin{equation}
{\Box}{\delta}a+m^{2}_{a,eff}{\delta}a=-g_{a\gamma\gamma}B_{0}\frac{m^{2}_{a,eff}}{{m^{2}}_{a}}{\partial}_{t}.
({\delta}A_{||})
\label{80}
\end{equation}
 Choosing the solution ansatz
\begin{equation}
{\delta}a(z,t)= e^{i\omega(z-t)}{\delta}a(z)
\label{81}
\end{equation}
and a similar expression for ${\delta}A_{||}(z,t)$, we obtain the following Schrödinger-like matrix equation
\begin{equation}
i\frac{d\psi}{dz}= A \psi,
\label{82}
\end{equation}
where the transpose of column matriz is ${\psi}^{T}=({\delta}a(z),{\delta}A_{||}(z))$,
and the matriz $A$ is given by
$\begin{pmatrix}
{\Delta}_{a} & {\Delta}_{a\gamma\gamma}\\
{\Delta}_{a\gamma\gamma} & 0\\
\end{pmatrix}$ 
where ${\Delta}_{a}$ and ${\Delta}_{a\gamma\gamma}$  are, respectively, the axion mass term and the mixing 
axion-photon term. They are explicitly given by
\begin{equation}
{\Delta}_{a} = \frac{m^{2}_{a,eff}}{2\omega},
\label{82}   
\end{equation}
and
\begin{equation}
{\Delta}_{a\gamma\gamma}=-\frac{1}{2}g_{a\gamma\gamma}B_{0}(\frac{m_{a,eff}}{m_{a}})^{2}.
\label{83}
\end{equation}
These expressions highlight the contributions of the torsion trace mass and the BI coefficient in photon-axion 
conversion mixing. The interplay of these factors is crucial for determining the system's dynamics.

This result has significant implications for fast radio bursts (FRBs) in astrophysics\,\cite{26}. The presence of the torsion 
trace mass and the BI coefficient can alter the interaction between photons and axions, potentially affecting the 
propagation and characteristics of FRBs. Understanding these effects could provide new insights into the fundamental 
physics governing these mysterious cosmic events. Further research in this area may lead to breakthroughs in our 
comprehension of the universe's dark matter and energy components, as well as the behavior of quantum fields under 
extreme conditions. The ongoing exploration of these phenomena promises to enrich our knowledge and may open new avenues 
for scientific discovery.

\section{Summary and Outlook}
This paper explores various classes of Proca gravity that incorporate torsion without metric-affine gravity, where 
torsion propagates in a vacuum. It demonstrates that, although the presence of ghosts often signifies new physics, 
when the spin-0 torsion mass is finite and Proca electrodynamics is not ghost-free, dark axion masses align with spin-0 
torsion masses through axion-driven torsion and photon-torsion mixing. This study reveals that Proca magnetogenesis with 
dark axions does not favor dynamo action but rather induces oscillations of magnetic fields via photon-torsion mixing 
for massless photons.

Recent studies have investigated dynamo and gravitational waves with massive photons in Einstein-Cartan-Holst gravity, 
showing minimal phenomenological effects due to the extremely weak photon mass in the Proca regime. Furthermore, it is 
concluded that photon-torsion mixing models with axial anomalies are free of ghosts and tachyons, making them suitable 
for quantum gravity applications. The paper also suggests that the inclusion of the Holst term in the Proca Lagrangian 
could lead to BI bounds.

In the future, exploring photon-torsion mixing with quantum corrections, such as those investigated by He et al.\,\cite{2i}, 
could unveil new facets of torsion dynamics, offering insights into dark matter and fundamental forces while refining our 
understanding of photon-torsion interactions. Additionally, photon-axion conversion presents intriguing contributions from 
the torsion mass spectrum. Investigating these could optimize axion-photon conversion conditions, enhancing detection and 
study of axions and impacting cosmological models.

Overall, our findings provide a strong foundation for further research into the complex relationships between torsion, 
axions, and photons. Continued exploration in these areas promises to deepen our understanding of quantum gravity and 
uncover new physics, broadening our comprehension of the universe. As we delve further, we anticipate making significant 
discoveries that will advance our knowledge and theoretical frameworks.

\section{Acknowledgements} I would like to express my gratitude to S Zell for discussions on Proca geometry and 
metric-affine gravity. Financial support from University of State of Rio de Janeiro (UERJ) is grateful acknowledged.
This work was performed under the auspices of Major Science and Technology Program of Xinjiang 
Uygur Autonomous Region through No.2022A03013-1.

\end{document}